\documentclass[nohyperref]{article}
\usepackage{microtype}
\usepackage{graphicx}
\usepackage{subfigure}
\usepackage{booktabs} 

\usepackage[nohyperref,accepted]{icml2022}

\usepackage{amsmath}
\usepackage{amssymb}
\usepackage{mathtools}
\usepackage{amsthm}
\usepackage{makecell}

\usepackage[capitalize,noabbrev]{cleveref}

\theoremstyle{plain}

\theoremstyle{definition}

\theoremstyle{remark}

\usepackage[textsize=tiny]{todonotes}

\begin{document}

\twocolumn[
\icmltitle{Target Geometry Estimation Using Deep Neural Networks in Sonar Sensing}



\icmlsetsymbol{equal}{*}

\begin{icmlauthorlist}

\icmlauthor{Chen Ming}{sch}
\icmlauthor{James A. Simmons}{sch}
\end{icmlauthorlist}

\icmlaffiliation{sch}{Department of Neuroscience, Carney Institute, Brown University, Providence, RI 02912}

\icmlcorrespondingauthor{Chen Ming}{gomingchen@gmail.com}

\icmlkeywords{biosonar, target geometry, glint spacing, echolocating bats, recurrent neural network, convolutional neural network}

\vskip 0.3in
]



\printAffiliationsAndNotice{}  

\begin{abstract}
Accurate imaging of target shape is a crucial aspect of wideband FM biosonar in echolocating bats, for which we have developed new algorithms that provide a solution for the shape of complicated targets in the computational domain. We use recurrent neural networks and convolutional neural networks to determine the number of glints (i.e., major reflecting surfaces) making up the target’s structure and the distances between the glints (target shape in sonar). Echoes are dechirped relative to broadcasts, and the dechirped spectrograms are scanned in short time segments to find local spectral ripple patterns arising from different interglint delay separations.  By proceeding in successive time-window slices, we mimic time-frequency neural processing in the bat’s auditory system as a novel means of real-time target discrimination for sonar sensing in robotics. 
\end{abstract}

\section{Introduction}
\label{intro}
There are around 1,000 species of bats that use echolocation in part or full for navigation and foraging~\cite{griffin1958listening,biologyofbats,1000species,hiryuglass}. The sonar pulses vary from constant frequency (CF) followed/preceded by a short frequency modulated (FM) signal, FM sweeps alone, clicks to noise bursts. The CF component is tuned to flutter detection from insects through Doppler shift compensation. The broadband FM signals are assumed to be used for range measurements, varying in duration, bandwidth, and repetition rate depending on the proximity to the prey~\cite{fenton1995natural,schnitzler2011auditory}. Big brown bats, \textit{Eptesicus fuscus}, use FM pulses spanning from 100 kHz to 20 kHz. They are one of the most widely studied bat species in
laboratory experiments~\cite{dear1993tonotopic,surlykke2000echolocation,hiryu2010fm,macias2018natural,amaroturnflight}. The sonar pulse duration can be less than 1 to tens of milliseconds. There are two acoustic cues in foraging: 1) the delay between emitted sonar pulse and the received echo determined through many frequency channels in the auditory information decoded by the cochlea, 2) the target geometry contained in interference patterns in spectrum from overlapping echoes reflected from surfaces within a complex target, such as insects. The two cues are decoded simultaneously in the auditory cortex by delay-tuned neurons and multi-peaked frequency-tuned neurons~\cite{dear1993tonotopic,simmons2012bats}. The first cue helps bats locate the prey, and the second provides information of geometry, size, and texture of the target~\cite{texture}. Both are critical information in echolocation. The interference patterns consist of repetitive spectral peaks and notches, and the frequency difference between neighboring notches/peaks is the reciprocal of the time delay of the glint spacing (GS), i.e., the elapsed time when sound travels from one glint (i.e., a reflecting surface) to another and then back. Thus, this frequency difference is the key to estimating target geometry~\cite{ming2021ploscomput}.  Behavioral experiments have also shown spectral notches are sufficient for bat to discriminate echoes from different parts of one target~\cite{simmons1990b}. Besides of the laboratory work, there are many studies focused on computational modeling of big brown bats, for example, sensorimotor models of foraging~\cite{kuc1994sensorimotor,mosssensorimotor} and jamming~\cite{yossisensorimotor}. There are also computational models of biosonar signal processing, such as ~\cite{ikuomodel} and~\cite{ming2021ploscomput}; the latter used parallel computational pathways for processing temporal and spectral information contained in the echoes and successfully reconstructed target shape using ripple patterns. The model combined the coarse- and fine-range estimation on the same perceptual axis. However, the limitation of using threshold tuning is that it extracts the profile of the time-frequency representation at the beginning, and thus only works for 2-glint targets that have well-defined and unique ripple patterns throughout the auditory spectrogram.

On the other hand, convolutional neural networks (CNN) have been widely used for human auditory research on sound localization~\cite{ma2017binaural,McdermontCNN} and for biosonar such as statistics of auditory tuning~\cite{sangwookbionet} in echolocating bats, natural landmark recognition~\cite{liujun}, and bat call detection~\cite{batcalldetectionCNN}. Recurrent neural network (RNN) and its subset long short-term memory (LSTM) have been used in sonar image classifications~\cite{perry2004rnn,yu2021rnn}.

In this paper, we will use ripple patterns to extract target geometry information, and extend the shape reconstruction to targets with three or more glints with RNN. We will also test the performance of a CNN and a RNN in a separate task - classifying the number of glints. The trained networks can be deployed on an autonomous vehicle for realtime target classification and identification using sonar sensing. The targets discussed in this paper are as big as insects, but can be scaled up for application in underwater sonars.


\section{Method}

\subsection{Biosonar Model}
The biosonar model simulating sonar emission and reception is inspired by big brown bats. The mouth of the bat is located at the origin, and two ears are at $(x,y,z) = (\pm 0.75, 0.75, 1.5$) cm. The mouth and two ears both have a radius of 0.5 cm, and are tilted $5^\circ$ down from the horizontal plane. Besides, the two ears are also tilted $25^\circ$ to the left and right sides. The detection sensitivity pattern of the circular ear aperture of radius $a_e$ at frequency $f$ has Bessel function form, thus the echo intensity at the inner ear can be calculated using~\cref{imr} assuming the prey is located at $(r_m, \beta_m)$ to the mouth and $(r_e, \beta_e)$ to an ear. The sampling frequency is 1 MHz.
\begin{equation}
\label{imr}
\begin{aligned}
    P_D(f,r_m,\beta_m,r_e,\beta_e) &= (\kappa\sigma\rho\pi U_0)\frac{fa_m^2a_e^2}{r_mr_e}\alpha(f,r_m+r_e)
    \\ 
    & \times (\frac{2J_1(ka_m\sin(\beta_m)}{ka_m\sin\beta_m}) 
    \\
    & \times (\frac{2J_1(ka_e\sin\beta_e)}{ka_e\sin\beta_e})\,,
\end{aligned}
\end{equation}

where $\kappa_e$ is a constant describing the geometrical properties of the pinnae, and $a_e^2$ is an indicator that hearing sensitivity improves with pinna size. The value of the constant term $(\kappa\sigma\rho\pi U_0)$ was set to 3000 for convenience. $\sigma$ is the scattering coefficient, equal to the ratio of backscattered pressure to incident pressure amplitudes ~\cite{kuc1994sensorimotor}. Although the biosonar model has two receivers, we only used one channel, or echoes received at the left ear, throughout this paper.

\begin{figure}[ht]
\begin{center}
\centerline{\includegraphics[width=1.1\columnwidth]{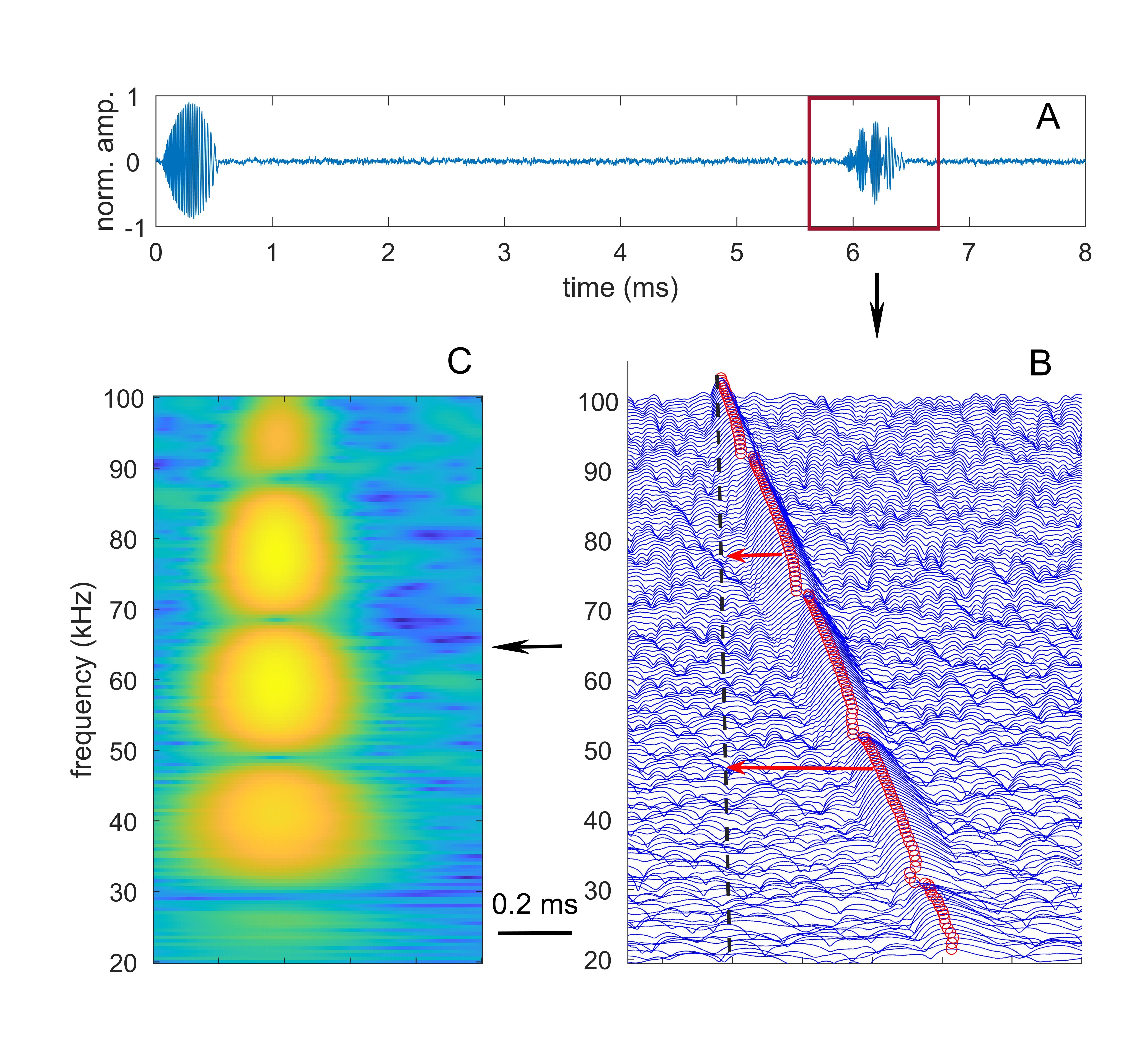}}
\caption{Calculation of gammatone spectrogram: A) The time series simulated by the sonar model, B) The waterfall plot of the gammatone filterbank output from the echo in maroon box in A). A universal threshold was used to acquire crossings on the broadcast and the echo, shown as the red circles. Those on the echo are corrected based on the crossings on the broadcast to maintain approximately the same delay between a pair of crossings on each frequency channel. The crossing at 100 kHz serves as the reference, and the echo was de-chirped by moving each frequency channel to the left by the number of samples that equals to the difference between local crossing and the reference, shown as the direction and length of the red arrows, C) The auditory spectrogram calculated from the dechirped echo. The values of the spectrogram were normalized to be within [-1,1]. The Yellow color indicates positive values, while blue color denotes negative values.}
\label{dechirp}
\end{center}
\vskip -0.3in
\end{figure}

The target was 1 m away from the mouth, and consisted of one to four glints. The first glint was always located at (0,1,0) m, and others were behind the first on y-axis. The biosonar model' outputs were broadcast-echo pairs. For example, a 2-glint echo was simulated in the model (see~\cref{dechirp}A). The first sound is the broadcast, and the following is the echo. The time delay between the broadcast and the echo is around 6 ms, which corresponds to the distance between sonar and target. Big brown bats emit FM sweep with two harmonics, first from 100 - 50 kHz and the second from 50 - 25 kHz~\cite{hiryu2010fm}. We chose a linear FM downsweep with one harmonic from 100 - 20 kHz for simplicity. Diffuse noise was added to make the model more realistic, and the signal-to-noise ratio was 20 dB~\cite{ma2017binaural}. 

\subsection{Auditory Spectrogram}
Gammatone filterbank is commonly used to mimic the propagation of sound along the basilar membrane (BM), a stiff structural element within the cochlea of the inner ear~\cite{yin2011gammatone,park2020gammatone}. At each point along the cochlea, the BM responds best to a certain frequency varying smoothly from high at the base to low at the apex. The movements of the BM cause the deflection of the hair cell which creates an electrical signal distributed throughout the rest of the auditory system~\cite{lyoncochleagram}. We used a variation called differentiated all-pole gammatone filter (DAPGF) ~\cite{gammatonevariation} to include the asymmetry observed in the tuning curves. The transfer function of the DAPGF filter is shown in~\cref{gamma}.

\begin{equation}
    H_\text{DAPGF}(s) = \frac{\omega_o^{2N-1}s}{[s^2 + \frac{\omega_o}{Q}s+\omega_o^2]^N}\,,
\label{gamma}
\end{equation}
where $N$ and $Q$ are the order and quality factor of the filter, and $\omega_o$ is the natural (or pole) frequency. Based on the tuning curves measured in the primary auditory cortex in big brown bats~\cite{tuningcurves}, we used $N=4$ and $Q=15$ through all frequencies from 20 kHz to 100 kHz with 0.5 kHz increment totaling 161 frequency channels. The calculation of de-chirped auditory spectrograms is shown in~\cref{dechirp}. The time series generated by the biosonar model (see~\cref{dechirp}A) was passed to the DAPGF filterbank, which produced the time-frequency representations (see~\cref{dechirp}B). To calculate the spectrogram, the filterbank output was de-chirped according to the highest frequency at 100 kHz by eliminating the number of samples between the threshold crossing and the black dotted line in~\cref{dechirp}B at the beginning of each channel, as if every channel was moved to the left to align the red circle with the black line. Then the spectrogram (see~\cref{dechirp}C) was acquired by calculating energy-by-band using a 0.128 ms window with 0.120 ms overlap. This provides a time bin of 0.008 ms in the auditory spectrogram, and a window of 250 time bins was selected to enclose the de-chirped echoes (see~\cref{dechirp}C). More auditory spectrograms of echoes with different numbers of glints are shown in~\cref{alltrainingdata}.

\subsection{Training Data}
\begin{figure}[ht]
\begin{center}
\centerline{\includegraphics[width=\columnwidth]{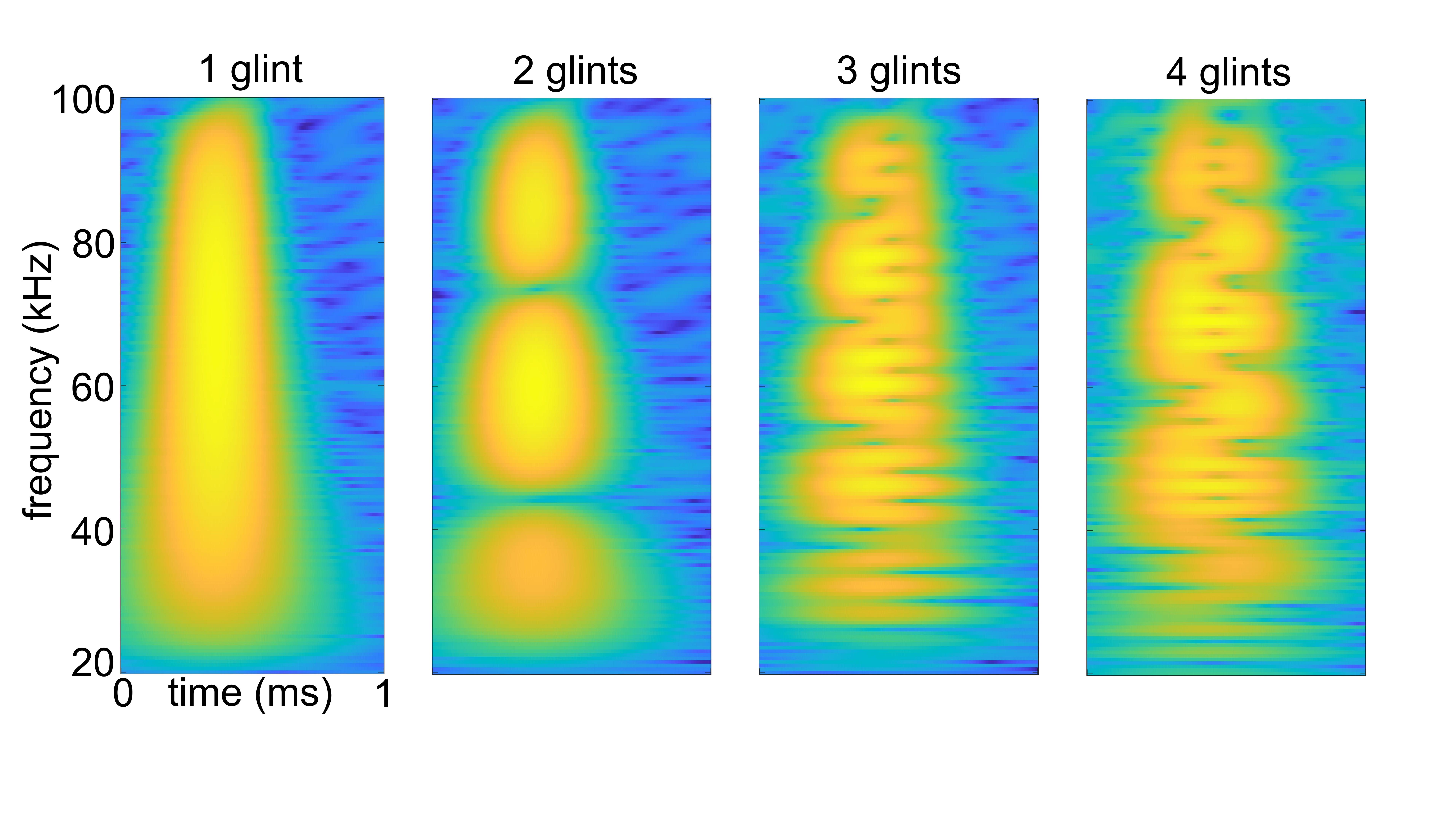}}
\caption{Examples of de-chirped spectrograms from targets with 1, 2, 3, and 4 glints, from left to right.}
\label{alltrainingdata}
\end{center}
\vskip -0.2in
\end{figure}

The training data were auditory spectrograms with a 2-ms-long window, or 250 time bins. The distance between the sonar and the target was always 1 m. A 2-ms window was demonstrated long enough to capture the entire de-chirped window regardless of the number of glints, shown in ~\cref{alltrainingdata}. There were 72 samples for each category. The 1-glint echoes were generated using broadcast with varying length from 0.5 ms to 10 ms; for targets with two glints, the glint spacings were 24 numbers linearly spaced between 3 mm to 7 cm; targets with 3 or 4 glints always had the first located at (0,0,0), and the rest located at random locations from 19 predetermined positions spanning linearly from 3 mm to 7 cm. Echolocating bats use pulses of various duration to survey the prey at different hunting stages~\cite{surlykke2000echolocation}, thus we used 0.5-, 3-, and 5-ms-long broadcasts for each category. The upper range limit 7 cm was chosen to account for various insect sizes. These resulted in 288 samples, which were split into train and test data for validation, with 56 samples assigned as test data. To train the LSTM model for glint spacing estimation, a total of 32 data samples were generated from 2-glint targets with linearly spaced distances between 0 and 7 cm. Five time bins were used for the parameter - timesteps - in the network. All training samples were normalized to be within [-1,1].

\subsection{Test Data for Evaluation}
To evaluate the performance of each trained network, we also generated new echoes with a different setup than the data used in training. Two signal duration of 0.7 and 4 ms were used, and 16 samples for each category were generated with similar approach as training data but with different numbers of points while generating the linearly spaced vector between 0 and 7 cm.

\subsection{Convolutional Neural Network}
\begin{figure}[ht]
\vskip 0.2in
\begin{center}
\centerline{\includegraphics[width=0.8\columnwidth]{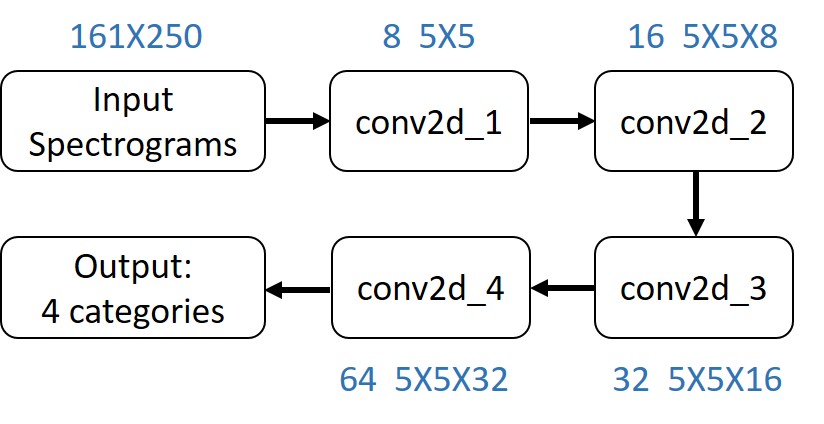}}
\caption{CNN architecture. The number of filters and filter dimensions are denoted in blue texts for each convolutional layer. There are 288 input spectrograms in total with a dimension of $161\times250$ each.}
\label{convarc}
\end{center}
\vskip -0.2in
\end{figure}

The spectrograms with dimension 161 $\times$ 250 (frequency channels $\times$ time bins) were passed to a convolutional neural network (CNN) that consists of 4 convolutional layers, each followed by a batch normalization, a relu activation layer, and a max-pooling layer. Those layers serve different purposes: 1) conv2d has a set of linear filters; 2) batch-normalization applied batch normalization to the output of the conv2d layer; 3) rectified linear units (ReLu) layer added pointwise nonlinearity; 4) max-pooling layer downsampled its input. The dropout layer randomly chose 50\% of the weights and set them to zero. The fully-connected layer, also often called a dense layer, flattened the input and applied weight and bias matrices for each input-output pair. The softmax and class output layers conducted the classification~\cite{McdermontCNN}. The number of filters in each convolutional layer is shown in~\cref{convarc}. The network was trained using the loss function:
\begin{equation}
\text{loss} = -\frac{1}{N}\sum\limits_{n=1}^N\sum\limits_{i=1}^K w_it_{ni}\ln{y_{ni}}\,,
\end{equation}
where $N$ is the number of samples, $K$ is the number of categories, $w_i$ is the weight for class $i$, $t_{ni}$ is the indicator that the $n$th sample belongs to the $i$th class, and $y_{ni}$ is the output for sample $n$ for class $i$.

\subsection{Recurrent Neural Network}
\subsubsection{Classification}
The Recurrent Neural Network (RNN) is a neural sequence model using sequential data, and are widely used in speech recognition~\cite{graves2014rnnspeech}, language modeling~\cite{mikolov2012rnnlanguage}, and machine translation~\cite{kalchbrenner2013rnntranslation}. Compared with feedforward neural networks, such as CNN, RNN has the memory of previous input in the sequence and uses that to influence the current output, where the previous inputs are represented in a hidden state within the network. Long short-term memory (LSTM) networks are a subset of RNN, which solves the problem of short-memory dependencies of a vanilla RNN network. In this paper, we used three LSTM layers, each of which has 128, 64, and 32 neurons with batch normalization and a dropout layer following the first two, shown in~\cref{rnnarc}. The dropout ratio is 0.3. The output shape indicates (batch size, timesteps, number of neurons). The batch size is 100. A variety of timesteps values are used to compare the performance, but here 5 time bins were shown in~\cref{rnnarc}.
The loss function is cross-entropy loss shown in~\cref{rnnloss} below:
\begin{equation}
\text{loss} = -\frac{1}{N}\sum\limits_{n=1}^N\sum\limits_{i=1}^K ((t_{ni}\log(y_{ni}) + (1-t_{ni})\log(1-y_{ni}))\,,
\label{rnnloss}
\end{equation}
where $N$ and $K$ are the numbers of observations and classes, respectively, and $w_i$ indicates the weight for class $i$.

\subsubsection{Glint Spacing Estimation}
Another RNN network was trained to distinguish the spacing between two glints. The timesteps value in the network was 5 time bins (0.04 ms), while big brown bats can discriminate time difference as small as 10 ns~\cite{simmons1990b}. The architecture is the same as the one used in the classification, but the training data for this network is 32 two-glint echoes with various spacing spanning from 0 - 7 cm, and was also shortened to 100 time bins (0.8 ms) from bin number 50 to 150 to eliminate the noise at two ends of the spectrogram. The output size of the network was also changed to 32 categories. When the glint spacing is 0 cm, the two glint echoes overlap at the same time delay, and thus have no spectral notches but appear like a single-glint echo (see~\cref{alltrainingdata}A). We count single-glint echoes as two-glint echoes with 0 glint spacing (ripple interval Inf) in this case. The trained network was then used to predict the glint spacing of 3- and 4-glint echoes by evaluating a 5-bin-long slice of the spectrogram sliding through. For example, the spectrogram (see~\cref{glintspace}A) was split into 20 time windows (see~\cref{glintspace}B). The estimates of glint spacing for each window are shown in~\cref{glintspace}C. Change points were found by detecting abrupt changes in the slope and intercept of the glint spacing estimates. The computational cost is linear in the number of observations\cite{changepointdetection}. Those points within 2 steps of other change points were eliminated due to that ripple pattern from one pair of glints usually lasts more than 5 steps, while noise combined with the incomplete spectrum at the beginning and the end can lead to frequent changes. The estimates following each verified change point indicate the predicted glint spacing values for a pair of glints.

\begin{table}[t]
\caption{RNN Architecture}
\label{rnnarc}
\vskip 0.15in
\begin{center}
\begin{small}
\begin{sc}
\begin{tabular}{lcccr}
\toprule
Layer & Type & Output Shape & Param \# \\
\midrule
1    & LSTM\_1               & (100, 5, 128) & 148480\\ \specialrule{.4pt}{0pt}{0pt}
2    & \makecell{batch- \\ normalization}  & (100, 5, 128) & 512   \\ \specialrule{.4pt}{0pt}{0pt}
3    & dropout(0.3)         & (100, 5, 128) & 0  \\ \specialrule{.4pt}{0pt}{0pt}
4    & LSTM\_2               & (100, 5, 64)  & 49408 \\ \specialrule{.4pt}{0pt}{0pt}
5    & \makecell{batch- \\ normalization}  & (100, 5, 64)  & 256       \\\specialrule{.4pt}{0pt}{0pt}
6    & dropout(0.3)         & (100, 5, 64)  & 0                         \\ \specialrule{.4pt}{0pt}{0pt}
7    & LSTM\_3               & (100, 5, 32)  & 12416  \\\specialrule{.4pt}{0pt}{0pt}
8    & flatten              & (100, 160)    & 0 \\\specialrule{.4pt}{0pt}{0pt}
9    & dense                & (100, 4)      & 644        \\
\bottomrule
\end{tabular}
\end{sc}
\end{small}
\end{center}
\vskip -0.1in
\end{table}

\section{Results}
\subsection{Glint Number Classification}
The classification performance of CNN and RNN is shown in~\cref{perf} with average values of final validation accuracy and training accuracy over 10 training sessions. The CNN can reach around 99.7\% validation accuracy after 400 epochs with a test accuracy of around 87\%. On the other hand, the RNN provides validation accuracy of 98.5\% and test accuracy around 83\%. The test accuracy is the percentage of correct predictions by the networks using 64 testing samples unseen during training and validation. The mistakes in predictions are mainly caused by the confusion between 3- and 4-glint targets, especially when the glints are clustered within a few millimeters. The timesteps used in RNN for classification in~\cref{perf} is 250 time bins, equivalent to the inputs for the CNN. The increase of epochs does not affect the accuracy of the CNN but boost the test accuracy of the RNN to 82.8\% from 75.8\%.

\begin{figure}[!htbp]

\begin{center}
\centerline{\includegraphics[width=1\columnwidth]{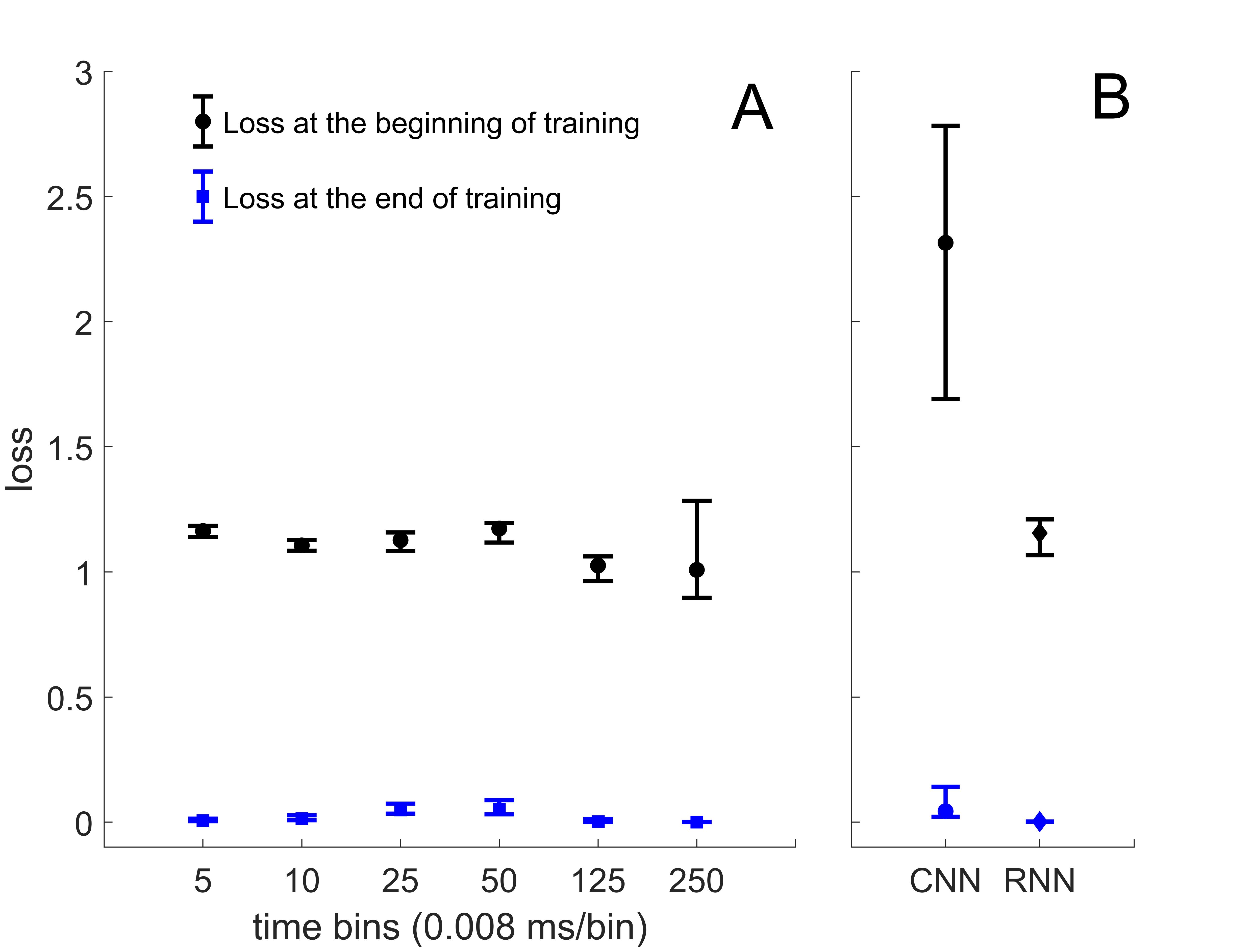}}
\caption{Network loss: A) Loss of the RNN over 10 networks with different timesteps inputs. The various timesteps values in training are from 5 bins to 250 bins. The losses at the beginning and end of the training are shown in black and blue colors, respectively. The filled circle and the square denote the median of losses over 10 networks. The up and down whiskers indicate the maximum and minimum value among the 10 training sessions, respectively. Batch size is 8 for all training sessions to accommodate different timesteps sizes, B) the loss of CNN and RNN over 10 networks with a batch size of 32, each trained with 100 epochs. The RNN has a timesteps input of 250 bins.}
\label{rnnvstimestep}
\end{center}
\vskip -0.2in
\end{figure}

As shown in~\cref{rnnvstimestep}A, the difference is not significant from using various time bin number as the timesteps input in RNN, but the performance of networks using 5, 10, 125, and 250 time bins is slightly better with a smaller final loss than the other two timesteps values. The comparison of initial loss and loss at the end of training between the two networks is shown in~\cref{rnnvstimestep}B. The CNN has higher starting and ending loss than the RNN network with the same epochs and input size. 

\begin{table}[t]
\caption{CNN and RNN performance}
\label{perf}
\vskip 0.15in
\begin{center}
\begin{small}
\begin{sc}
\begin{tabular}{lccccr}
\toprule
         & \multicolumn{2}{c}{CNN}             & \multicolumn{2}{c}{RNN} \\
\midrule

epoch    & 100      & 400            & 100  & 400 \\
\specialrule{.4pt}{0pt}{0pt}
\makecell[l]{validation \\ accuracy (\%)}  & 96.1 & 99.7  & 96.3  & 98.5 \\
\specialrule{.4pt}{0pt}{0pt}
\makecell[l]{final \\ accuracy (\%)}    & 99.4    & 100  & 100    &  100\\
\specialrule{.4pt}{0pt}{0pt}
\makecell[l]{test \\ accuracy (\%)}     & 86.9    &  85  &  75.8  &     82.8   \\
\bottomrule
\end{tabular}
\end{sc}
\end{small}
\end{center}
\vskip -0.1in
\end{table}


\subsection{Glint Spacing Estimation}
\begin{figure}[ht]
\begin{center}
\centerline{\includegraphics[width=0.8\columnwidth]{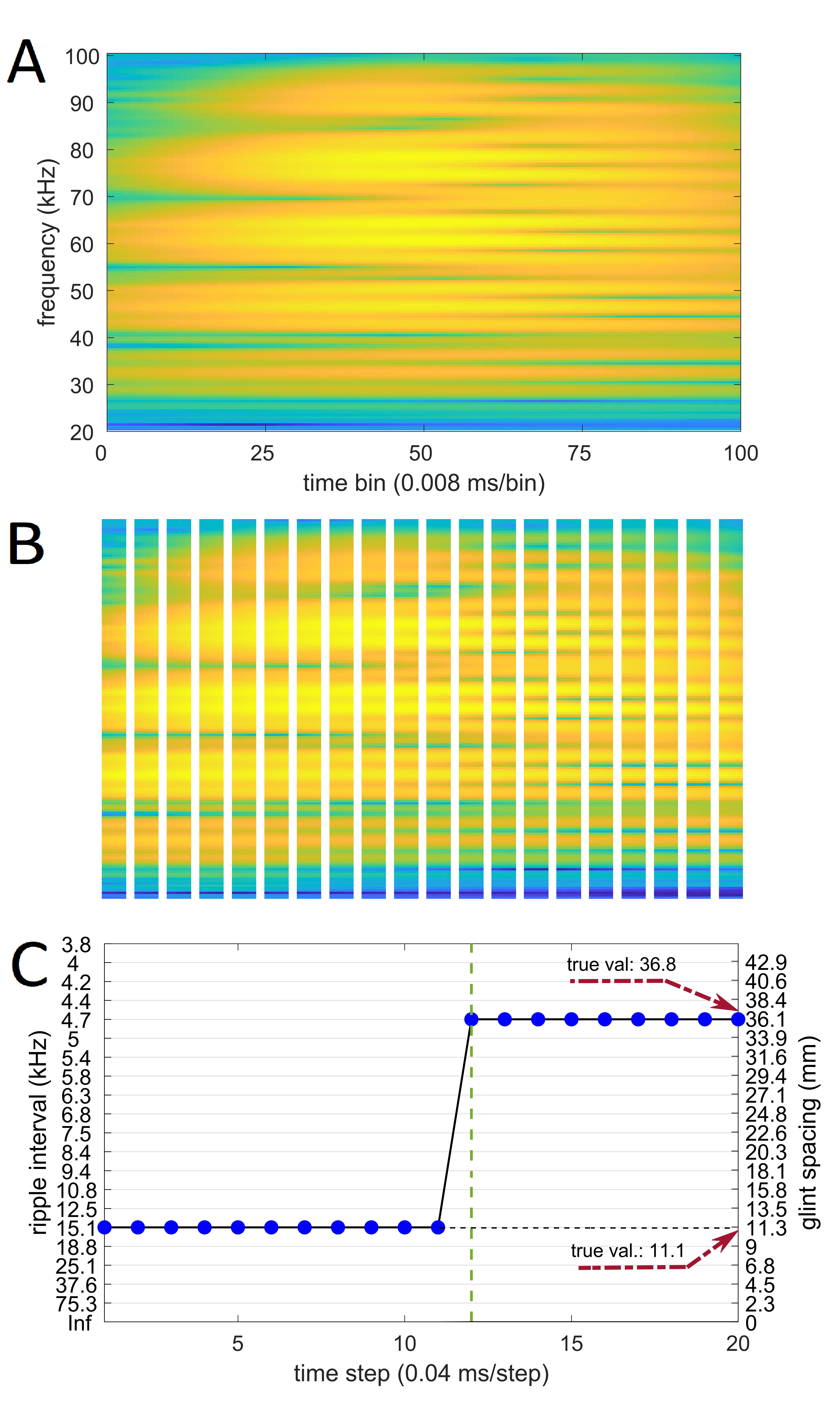}}
\caption{Glint spacing estimates for a 3-glint echo: A) the spectrogram cut to a shorter time window - 0.8 ms, to eliminate irrelevant noise. The GS values are 11.1 mm at the beginning and then 36.8 mm at the end of the spectrogram, B) The spectrogram was split into 20 windows. Each has the dimension of $161\times5$, C) The predictions of glint spacing by the trained network. The left y-axis indicates the frequency interval between neighboring spectral peaks/notches, while the right y-axis shows the corresponding distance between the two glints. The true values of the glint spacing were marked using red dashed arrows. The green dashed line depicts the changing point in the estimations.}
\label{glintspace}
\end{center}
\vskip -0.2in
\end{figure}

\begin{figure}[ht]
\begin{center}
\centerline{\includegraphics[width=\columnwidth]{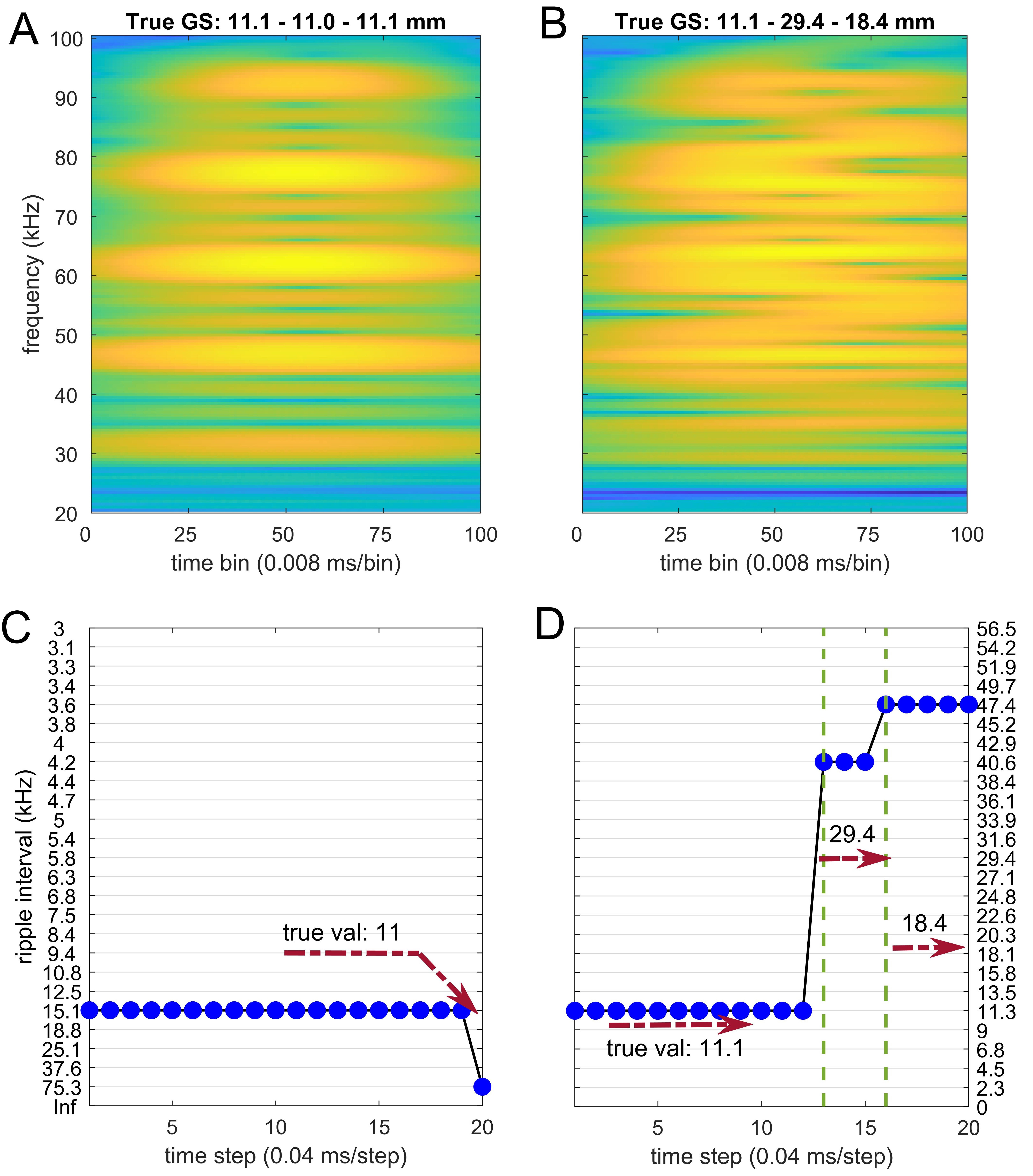}}
\caption{Glint spacing estimates for 4-glint echoes: A) spectrogram of a 4-glint echo with 11 mm glint spacing between each pair of glints, which means the locations of the four glints are (0,0,0), (0,11,0), (0,22,0), (0,33,0) mm, B) spectrogram of a 4-glint echo with glint spacings equal to 11, 30, and 18 mm, C) the glint spacing estimate for A), D) the glint spacing estimate for echo in B). The true values of GS at each section of the spectrograms are also marked with red arrows.}
\label{fourglintest}
\end{center}
\vskip -0.2in
\end{figure}

The average final loss and accuracy are around 0.029 and 99.6\% over 10 nets, respectively. A 3-glint echo is shown in~\cref{glintspace}A, where the distance between the 1st and 2nd glint is 11.1 mm, and that between the 2nd and the 3rd glint is 36.8 mm. The estimates of ripple interval for each window in~\cref{glintspace} were plotted in~\cref{glintspace}C. They are closest to the label that best approximates the true glint spacing. The change point was detected in the estimates array and its location was shown in a green dashed line in ~\cref{glintspace}C. Each change point indicates a start of consistent glint spacing estimates that last more than 5 steps. Because this change point appears in the middle, it demonstrates that there is another set of glint spacing estimates before, i.e., the target has three glints. Another example~\cref{fourglintest} shows the glint spacing estimation of two 4-glint echoes. While the estimates for~\cref{fourglintest}A are accurate (shown in~\cref{fourglintest}C), the change point was not detected due to a uniform spacing between neighboring glints. Estimates of~\cref{fourglintest}B only predicted the glint spacing of the first pair correctly (shown in~\cref{fourglintest}D).

\section{Discussion}
Researchers have shown the signatures in echoes contain crucial information for the classification of different types of targets, such as trees of different species~\cite{yovel2008plant,ming2017foliageclassification,ming2017simplified}, and targets like sphere, cube, and cylinder~\cite{mossTarget1995,sutlive2019echosignature}. However, those studies did not quantify the size and geometry of those targets from echoes, which are important yet challenging tasks for underwater sonar systems~\cite{hodges2011underwater}. In this paper, we have filled in the gap by successfully reconstructing complex targets with three and more major reflecting parts by using ripple patterns of the time-frequency representation in the auditory information relayed from cochlea to the auditory neural pathways. This paper has also demonstrated neural processing in the auditory system can instruct the architecture design of artificial neural networks for target information decoding. Specifically, we used RNN for the geometry estimation of targets with three and more glints, compared with a previous model~\cite{ming2021ploscomput}. We showed RNN can be trained to recognize ripple patterns, the representation that multi-peaked frequency-tuned neurons in the auditory cortex are sensitive to for the decoding of prey size and target texture. The glint spacing estimation combined with change point detection is proven to be effective in target geometry reconstruction for 3-glint targets. We have also shown that CNN and RNN are effective in glint number estimation. The two networks can classify the number of glints with good accuracy. The CNN performs slightly better than the RNN in testing unseen data, although a more thorough comparison needs to be done by varying the architecture and tuning hyperparameters. As for the RNN, the length of the parameter - timesteps - affects the initial and final losses, which suggests a preference for short or longer windows.

The gammatone spectrograms of 2-glint echoes were split into 0.04-ms-long time windows and used as training data for the target shape reconstruction. The trained RNN can estimate the time window of the same length in 3- and 4-glint echoes. The window is short enough to exclusively show the spectral notches and peaks from a single pair of glints, which reduces the complexity of the task. However, 4-glint echoes exhibit more complicated ripple patterns due to the coupling from multiple glint echoes, so the prediction is less accurate. From a series of glint spacing estimates, we detected change points to group them for geometry reconstruction. As we have observed during testing, the missing high frequencies could affect the prediction accuracy (see the last estimate in~\cref{fourglintest}C), which aligns with the discovery in behavioral experiments~\cite{bates2011science} that big brown bats use the full spectrum to distinguish the targets from surroundings. An echo with missing highest few frequencies may be perceived as clutter echo, since the acoustic beam of biosonar has narrower beamwidth at high frequencies and by aiming at targets of interest, off-axis clutter only receives low-frequency content. Bats also utilize the tail-end frequency to match the incoming echoes with the right calls in a crowded environment~\cite{hiryu2010fm,mingpnas}. Those are also the reasons why cross correlation may not work for broadband signals in biosonar sensing. The absence of just a few kilohertz from 80 kHz will still create a strong peak, suggesting high correlation between broadcast and the echo, though the echo should be discarded instead. This paper used Welch window~\cite{welchwindow} on the call to expand the applicability to robotics, since signals with sharp transitions at two ends will cause spectral leakage in acquired digitized signals.

The targets discussed in this paper consist of glints lined up along y-axis with an assumption that there is no shielding effect. They are a simplification of insects. However, an ensonification of any target will come down to echoes of the glints at the front of the target because high frequency waves have smaller wavelength than the target, and the distance between the sonar and each glint is reflected in the echoes as time delay~\cite{moss1994mothecho}, or the coordinate on the axis which the sonar is facing (y-axis in this paper). So the simplification is inspired from the acoustics aspect and does not affect the models' applicability to real-life targets. On the other hand, the azimuth information of a specific glint can be inferred by binaural estimation.


Besides, we have tried using time series of the signal and a RNN to identify the notches visible in the signal. While 2-glint echoes have repetitive up-down patterns in the time series, adding one more glint will disrupt the pattern. On the other hand, the gammatone spectrogram expands the information from time series to 161 frequency channels, which enables the decoding of information that the time series could not. The CNN has been demonstrated to have good performance in distinguishing echoes with different numbers of glints, yet it cannot decode the target geometry with more than two glints. Previous trials with a CNN using spectrogram and labels of regions of interest (ROI) around the spectral notches were not successful. The reason could be that those dark spots (notches) on a spectrogram are not distinct from other features, such as places having a lower amplitude due to the cancellation by noise. This highlights the importance of ripple patterns - a slice of spectrogram along the frequency direction - in determining the target structure. 

Since biosonar is very efficient and accurate, there has been a lot of interest in building bat-inspired sonar sensing robots~\cite{philipterrain,hiryurobot,rolfbatbot}, which used the bat calls alike signals or even included the pinnae movement of horseshoe bats. Sonar sensing works well in low light conditions, and thus can be applied to autonomous drones in tasks such as surveying the regions with fire and delivery at night. As the small-footprint mobile deep learning gets more attention recently~\cite{smallfootprint2015, squeezenet2016, smallfootprint2021}, sonar sensing will have more impact in small autonomous vehicles. We aim to design an autonomous vehicle equipped with a parsimonious model with reduced number of parameters in the near future. Though small targets are the focus of this paper, the target size can be scaled up with lower frequencies for potential applications in Navy sonar.




\bibliography{ming_submission}
\bibliographystyle{icml2022}

\end{document}